\documentclass[twocolumn, twocolappendix]{aastex631}

\shorttitle{Ultrashort-period WD binaries: no strong tidal heating}
\shortauthors{Scherbak and Fuller}

\begin{document}

\title{Ultrashort-period WD binaries are not undergoing strong tidal heating}

\author{Peter Scherbak}
\affiliation{California Institute of Technology, Astronomy Department, Pasadena, CA 91125, USA}

\author{Jim Fuller}
\affiliation{TAPIR, California Institute of Technology, Pasadena, CA 91125, USA}




\begin{abstract}

Double white dwarf (WD) binaries are increasingly being discovered at short orbital periods where strong tidal effects and significant tidal heating signatures may occur. We assume the tidal potential of the companion excites outgoing gravity waves within the WD primary, the dissipation of which leads to an increase in the WD's surface temperature. We compute the excitation and dissipation of the waves in cooling WD models in evolving \texttt{MESA} binary simulations. Tidal heating is self-consistently computed and added to the models at every time step. As a binary inspirals to orbital periods less than $\sim$20 minutes, the WD's behavior changes from cooling to heating, with temperature enhancements that can exceed 10,000 K compared with non-tidally heated models. We compare a grid of tidally heated WD models to observed short-period systems with hot WD primaries. While tidal heating affects their $T_{\rm eff}$, it is likely not the dominant luminosity. Instead these WDs are probably intrinsically young and hot, implying the binaries formed at short orbital periods. The binaries are consistent with undergoing common envelope evolution with a somewhat low efficiency $\alpha_{\rm CE}$. We delineate the parameter space where the traveling wave assumption is most valid, noting that it breaks down for WDs that cool sufficiently, where standing waves may instead be formed. 

\end{abstract}



\section{Introduction}

Double white dwarf (DWD) binaries have increasingly been found at short orbital periods (e.g., \citealt{brown_12_2011, burdge_systematic_2020}), with the tightest detached system at an orbital period of only 6.9 minutes \citep{burdge_general_2019}. For these compact systems, tides are expected to be very strong (e.g. \citealt{iben_luminosity_1998}), but the effects of tides on the WDs and the resulting observable predictions have remained uncertain.

As DWD binaries inspiral to shorter orbital separations due to the emission of gravitational waves (GWs), tidal effects will become stronger, and tidal heating could become the dominant source of luminosity \citep{iben_luminosity_1998, burdge_88_2020}. Measuring tidal heating is one of the principle ways to constrain tidal physics \citep{piro_inferring_2019}. At very short periods, tides may affect the pre-merger conditions of DWD binaries, such as the synchronization of WD rotation with the binary orbit. The degree of synchronization pre-merger is unclear, with some simulations assuming the WDs are tidally locked (e.g., \citealt{raskin_remnants_2012, dan_structure_2014}), whereas others have assumed the WDs are initially non-rotating (e.g., \citealt{yoon_remnant_2007, loren-aguilar_high-resolution_2009}). When comparing WD merger simulations with these two different initial conditions, \cite{dan_structure_2014} found that the merger timescale and the location of helium detonation were significantly altered. Therefore, the degree of synchronization seems to affect the merger product of two WDs, which is a pathway to Type Ia supernova (e.g., \citealt{shen_every_2015}) and other exotic phenomena.

While tides are expected to contribute to heating of both WDs, quantitative predictions of the heating, and its dependence on WD mass and system orbital period, have not been rigorously performed. Instead, the magnitude of the tidal heating has only been estimated for a few different WD masses and orbital periods \citep{fuller_dynamical_2013, yu_tidally_2021}, and the changing WD structure, heating rate, and location of heat deposition have not been self-consistently modeled. Many recently discovered short-period DWD binaries have measured masses and surface temperatures, meaning that precise predictions of tidal heating can now be directly compared to observations.

In a compact binary, each component feels a tidal force due to its companion's gravitational potential. The equilibrium tidal response, referring to a body's overall deformation in quasistatic equilibrium, is unlikely to be important except when the binary is close to mass-transferring \citep{fuller_tidal_2011}. Rather, the wave-like or oscillatory tidal response, known as the dynamical tide \citep{zahn_dynamical_1975, zahn_tidal_1977}, is likely dominant. The dynamical tide often involves the excitation of standing-wave gravity modes \citep{zahn_tidal_2008}, or, if there is sufficient dissipation, a train of traveling gravity waves  \citep{goldreich_tidal_1989}. In  stars on the main sequence, waves are excited at the convective-radiative boundary and their dissipation can drive circularization and synchronization of the binary.
Gravity modes are also found in WDs, as demonstrated observationally by pulsating ZZ Ceti stars \citep{brickhill_pulsations_1983}.
The nature of the tidal excitation, and whether it is in the standing wave or traveling wave regime, will help determine the impact of tides on a WD's observed properties. Previous works that have focused on standing waves in WDs include \cite{burkart_tidal_2013} and \cite{yu_non-linear_2020}.

In a series of papers, \cite{fuller_tidal_2011}, \cite{fuller_dynamical_2012}, \cite{fuller_dynamical_2013} applied the theory of dynamical tides to a DWD binary. In particular, \cite{fuller_dynamical_2012} and \cite{fuller_dynamical_2013} found that, for both CO- and He-core WDs, excited gravity modes are likely to damp in the outer envelope and prevent the formation of standing waves. 
\cite{fuller_dynamical_2013} modeled tidal heating in a WD model, but this analysis was limited to a WD model of fixed mass initialized at fixed temperatures. 

In this paper, we  investigate tidal effects for WDs of various masses that undergo evolution in a binary and are allowed to realistically cool (if tidal heating is unimportant) or be heated by tides. Only with a grid of models can simulations be compared to known systems to better interpret observations.
In addition, \cite{fuller_dynamical_2013} assumed that the tidal response is in the strongly dissipated traveling wave regime, and focused on nonlinear wave breaking as the most important dissipative mechanism. Here we model the contributions of both nonlinear wave breaking and radiative damping as dissipative mechanisms, and estimate where the overall traveling wave assumption breaks down.

In this work, we self-consistently implement tidal effects into cooling He-core WD models simulated in the \texttt{MESA} stellar evolutionary code \citep{Paxton2011, Paxton2013, Paxton2015, Paxton2018, Paxton2019}. Sec. \ref{Tide basics} reviews the physics of the excitation of gravity modes/gravity waves in a DWD binary. Sec. \ref{effects of tides} describes the implementation of dynamical tides in a DWD binary \texttt{MESA} simulation, where the WDs undergo spin-up and heating due to the dissipation of excited gravity waves. Sec. \ref{heating} presents the models where the traveling wave response works best and where it breaks down.  Sec. \ref{Tidal heating compare} presents tidal heating predictions for simulations of differing WD masses and initial orbital periods, and compares these predictions to observed ultrashort-period DWD binaries. We discuss our results in Sec. \ref{discussion} and conclude in Sec. \ref{conclusion}.

\section{Basic tidal physics}

\label{Tide basics}

More detailed analysis of tidally excited oscillations in WDs is found in \cite{fuller_dynamical_2012}. Here we review the basic quantities needed for our analysis. 

Consider a DWD binary with primary mass $M_1$ and secondary mass $M_2$ in a circular orbit of separation $a$ and orbital frequency $\Omega$. The following analysis treats the secondary as a point mass and only models tidal effects in the primary $M_1$, which has a radius $R$. 
The presence of $M_2$ creates a tidal potential that acts as a forcing term in the fluid equations describing the primary's interior. The tidal potential is predominantly quadrupolar with degree $l=m=2$. It varies sinusoidally in time with tidal frequency 
\begin{equation}  \label{tidal freq}  
    \omega = 2(\Omega-\Omega_s),
\end{equation}
\noindent where $\Omega_s$ is the spin frequency of the WD.


In the outer region of the WD, the tidal potential can either excite standing g-modes (with a reflective outer boundary condition) or gravity waves (with an outgoing boundary condition). In either case, the restoring force for these oscillations is buoyancy. The Brunt–Väisälä  frequency $N$, or buoyancy frequency, is therefore very important in determining where these modes/waves are excited and where they can propagate.

Because of strong dissipative effects, the wave energy may be depleted before the wave is reflected. Therefore, in this work we assume the response is in the outgoing wave regime (but see Sec. \ref{heating} and Figs. \ref{fig:wave regime 0.25} - \ref{fig:wave regime 0.45} for when this assumption breaks down). For a He-core WD, the outgoing wave is excited near the composition gradient at the transition between the WD's He core and H envelope \citep{fuller_dynamical_2013}, where there is a spike in $N^2$.

In the WKB limit, gravity waves satisfy the dispersion relation

\begin{equation}\label{dispersion}
    k_{r,\rm{WKB}}^2 = \frac{ \left(L_l^2 - \omega^2 \right) \left(N^2 - \omega^2 \right)  }{a_s^2 \omega^2} 
\end{equation}
\noindent with the Lamb frequency
\begin{equation}\label{lamb}
  L_l^2 = \frac{l(l+1) a_s^2}{r^2},
\end{equation}
\noindent where $k_r$ is the radial wavenumber, $a_s$ is the adiabatic sound speed, and $r$ is the internal radius of the WD. We leave $l$ general in our equations although $l=2$ is always assumed in this paper.


In most of the WD's interior, the gravity waves satisfy $\omega^2 \ll L_l^2$ and $\omega^2 \ll N^2$. However, there are regions near the surfaces of our WD models where $\omega^2$ becomes comparable to $L_l^2$  (see also Sec. \ref{wave regime sec} for how the traveling wave approximation can break down in these cases). If applied to Eq. \ref{dispersion}, $k_r^2$ would become very small. We therefore assume a lower limit of $k_r^2$ as $1/\left(4 H^2\right)$, where $H$ is the pressure scale height. Therefore, our full expression for the radial wave number is
\begin{equation} \label{dispersion full}
     k_r^2 = max \left[ \left(L_l^2 - \omega^2 \right) \left(N^2 - \omega^2 \right) / a_s^2 \omega^2, \frac{1}{4 H^2} \right].
\end{equation}

The wave carries both an angular momentum (AM) flux and energy flux toward the surface of the WD. AM and energy will therefore be transferred from the binary orbit to the WD, causing the WD to undergo spin-up and heating.

\section{Tidal effects in MESA simulations}

We create He-core WDs in the \texttt{MESA} stellar evolutionary code \citep{Paxton2011, Paxton2013, Paxton2015, Paxton2018, Paxton2019}. The WD models are created by mass stripping of an evolved red giant branch star when its He core reaches masses of 0.25, 0.30, 0.35, 0.40, and 0.45 $M_\odot$, giving 5 WD masses in our grid. We varied the mass of hydrogen in these WD models from about $10^{-5} $ to $10^{-3} $ $M_\odot$, but found that our tidal spin-up and tidal heating results were ultimately not greatly affected, with parameters like WD mass and age mattering much more (see Appendix \ref{H effect sec} for further discussion). However, if the hydrogen envelope was completely stripped, we would not expect a spike in Brunt–Väisälä frequency where the gravity waves are excited and our assumptions would greatly break down. Therefore, our results likely only apply to DA WDs with a hydrogen atmosphere.

We simulate tidal interactions with the WD in a binary \texttt{MESA} simulation with a point mass companion. The WD is initialized at a hot surface temperature post-mass stripping, meant to represent a newly-formed WD at the top of its cooling track. We therefore model tidal effects in the younger WD, assuming the companion WD already exists. If tidal heating is not implemented, the WD cools monotonically. The binary system evolves only due to the emission of GWs. The tidal backreaction of tides on the orbit is not implemented, because it is subdominant \citep{fuller_dynamical_2012, burdge_general_2019}. We initialize the binaries at a range of initial orbital periods, which represents the birth period of the DWD binary, i.e. when the younger white dwarf formed. It also corresponds to the post-CE period of the binary, as these DWDs likely went through common envelope evolution (e.g. \citealt{nelemans_reconstructing_2005}).

\label{effects of tides}


\subsection{WD spin-up}

\label{spinup}

\begin{figure}
    \centering
    \includegraphics[scale=0.6]{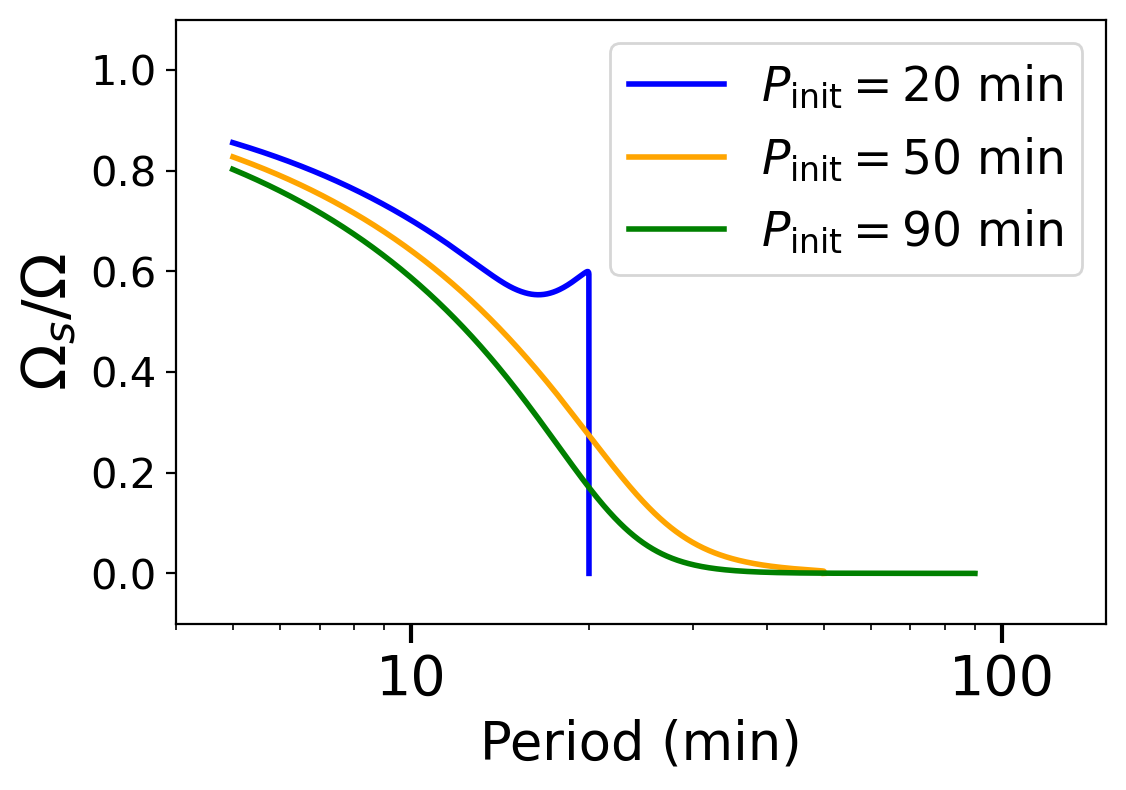}
    \caption{Evolution of the spin frequency $\Omega_s$ in units of the orbital frequency $\Omega$ as a function of the orbital period, due to tidal synchronization. The WDs start with $\Omega_s$ = 0. The WD mass is fixed at $0.35 M_\odot$ and WD models start evolution at the same temperature, but in a binary with different initial orbital period. The companion mass is fixed at $0.3 \ M_\odot$.}
    \label{fig:spinup}
\end{figure}

We assume that the WD is rigidly rotating and therefore has a constant $\Omega_s$ from surface to core (the validity of this assumption is discussed in Appendix \ref{AM transport}). \cite{fuller_dynamical_2013} finds that the tidal torque $\dot{J_z}$ can be written as 
\begin{equation} \label{torque}
     \dot{J_z} = T_0 F(\omega).
\end{equation}
\noindent Here, $T_0 = G M_2^2 R^5/a^6$ and represents how the magnitude of the tidal forcing scales with the binary system parameters. $F(\omega)$ quantifies the strength of the excited wave and is determined by solving for the waveform of tidally-excited gravity waves at various frequencies. The numerical solver developed in \cite{fuller_dynamical_2013} calculates the waveform by solving the equations of stellar oscillation with a tidal forcing term, assuming an outgoing boundary condition for the wave. For our WD models of varying mass, we calibrated $F(\omega)$ by running the solver on \texttt{MESA} profiles saved at various ages. We found that $F(\omega)$ approximately scales as 
\begin{equation} \label{f_omega}
    F(\omega) = \hat{f} \hat{\omega}^7
\end{equation}
\noindent where $\hat{\omega}$ is $\omega$ in units of $\sqrt{GM/R^3}$ and $\hat{f}$ is a function of WD mass and age. This power-law scaling is slightly steeper than that found in \cite{fuller_dynamical_2013}, where $ F(\omega) = \hat{f} \hat{\omega}^6$. However, the values of $\hat{f}$ are similar, and increase as the WD ages. We implement an interpolation of $\hat{f}$ into \texttt{MESA}, interpolating as a function of WD mass and age.
As a rough estimate, $\hat{f} \approx \hat{f_0}  t_{\rm age}^{0.4}$ where $t_{\rm age}$ is the age of the WD 
up to $\approx 1$ Gyr and $\hat{f_0} = 3 \times 10^{-6}$ for a 0.35 $M_\odot$ 
WD. $\hat{f_0}$ is larger (smaller) by a factor of $\approx 3$ for a 0.45 (0.25) $M_\odot$ WD.




Given the tidal torque, the spin-up rate of $\Omega_s$ is
\begin{equation} \label{omegasdot}
     \dot{\Omega}_s = \frac{\dot{J_z}}{I} = \frac{ T_0 \hat{f}  \hat{\omega}^7}{I}
\end{equation}
\noindent where $I$ is the moment of inertia of the WD. 
Eqs. \ref{tidal freq} and \ref{omegasdot} are coupled and can be numerically solved for $\Omega_s$ and $\omega$  during the \texttt{MESA} simulation. We assume the WD starts with $\Omega_s = 0$ at some orbital frequency $\Omega$, then allow the system to evolve. The WD cools, and the binary inspirals to closer separations/higher orbital frequencies due to the emission of GWs. 

The tidal spin-up is shown in Fig. \ref{fig:spinup} for a WD of 0.35 $M_\odot$ starting at orbital periods of 20, 50 and 90 min. The WD's spin becomes synchronized with the orbit at short orbital periods ($\Omega_s/\Omega \rightarrow 1$). By the time the binaries are at periods $\lesssim 30$ min, the binaries starting at 50 and 90 min are in the synchronizing regime where $\dot{\Omega}_s \approx \dot{\Omega}$. The WD starting in a binary at 20 min spins up very rapidly at first because the tidal torque is strong and it is in the short-period regime where synchronization would be expected. However, the young WD's radius also contracts rapidly, leading temporarily to a decreased torque and a dip in the value of $\Omega_s/\Omega$, before continuing to approach synchronization at shorter periods.

We find similar behavior for He-core WDs of masses 0.25 to 0.45 $M_\odot$, where the spin becomes increasingly synchronized with the orbit at orbital periods $\lesssim$ 30 minutes.

\subsection{WD heating}

\label{heating}

        

With the evolution of spin frequency $\Omega_s$ and tidal frequency $\omega$ known, the tidal heating can also be estimated. There are two basic steps to the process: \textbf{(1)} Calculate the magnitude of the power carried by the outgoing gravity wave, $L_{\rm tide}$. \textbf{(2)} Calculate the energy dissipated as heat (therefore lost from the wave) at each mass coordinate $m$ within the WD, between the location of wave excitation and the upper boundary of the wave propagation cavity. The heat is then injected into the WD model via the \texttt{MESA} control \texttt{s\% extra\_heat} (with dimension energy/time/mass).
The power $L_{\rm tide}$ carried by the wave in the corotating frame of the WD (for the dominant $m$ = 2 component of the tidal response) is
\begin{equation}
    L_{\rm tide} = \frac{\omega}{2} \dot{J_z} = \frac{\omega}{2} T_0 F(\omega) \simeq \frac{\omega}{2} T_0 \hat{f}  \hat{\omega}^7.
\end{equation}


\begin{figure*}
    \centering
    \includegraphics[scale=0.6]{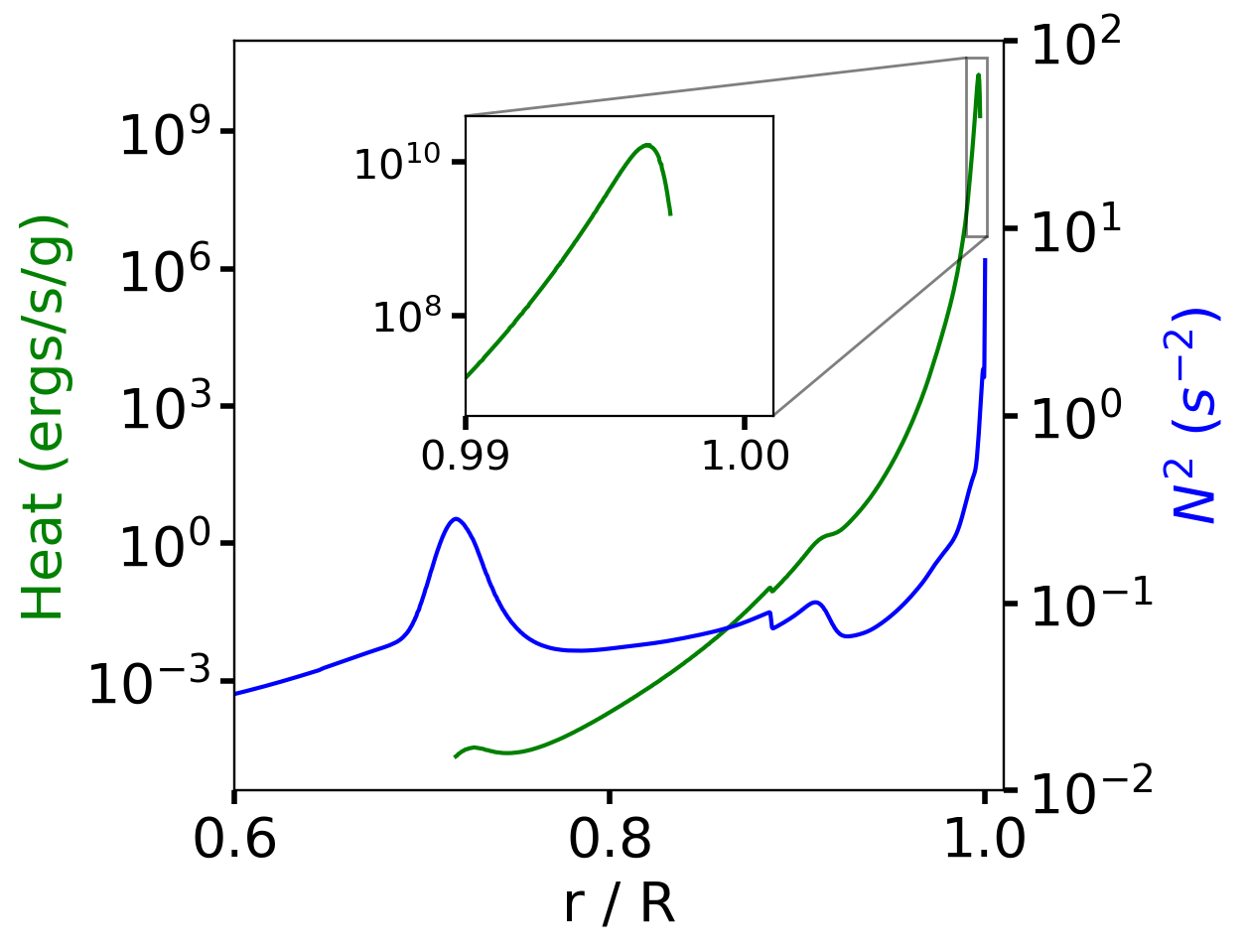}
        \caption{ Gravity wave excitation in the outer region of a 0.35 $M_\odot$ WD, plotted versus the radius $r$. The outgoing wave is assumed to originate at the bump in the Brunt–Väisälä frequency squared $N^2$ (blue curve) corresponding to the core-envelope composition gradient.  Heating due to wave dissipation (green curve) only takes place at exterior locations to the composition gradient. This snapshot is shown for an orbital period of 14.3 minutes, and a companion mass of 0.3 $M_\odot$.  
        }
        
    \label{fig:internal_wave}
\end{figure*}

The next step is to calculate the energy dissipated as heat in each cell of the model as the traveling wave propagates outward. In practice, the gravity wave never reaches the surface of the WD because its energy is either dissipated as heat or the wave reaches the end of its propagation cavity, where either $\omega^2 > L_l^2$ or $\omega^2 > N^2$. Moving outward from the excitation location, energy is removed from the wave via two dissipative mechanisms: nonlinear wave breaking and damping due to radiative diffusion. The decrease in $L_{\rm tide}$ over over a cell of mass $dm$, $d L_{\rm tide}/ dm$, is the power per unit mass deposited in a cell, and thus the necessary value for \texttt{s\% extra\_heat}. 

If radiative damping alone is included,

\begin{equation}
    \frac{d L_{\rm tide}}{ dm } = -\frac{L_{\rm tide}}{m_{\rm damp}} \, ,
\end{equation}
\noindent where $m_{\rm damp}$ is the characteristic mass scale for damping. This is related to the wave damping length scale $h_{\rm damp}$ by 

\begin{equation}
    m_{\rm damp} = 4 \pi \rho r^2 h_{\rm damp}
\end{equation}

\noindent and the length scale for thermal diffusion is 

\begin{equation} \label{damping}
    h_{\rm damp} = v_g/\left(k_r^2 K\right)
\end{equation}

\noindent where $v_g \approx \omega/k_r$ is the group velocity of the wave, $\rho$ is the mass density, and $K$ is the thermal diffusivity (e.g. \citealt{fuller_pre-supernova_2017}). $K$ is given by 

 \begin{equation}
     K = \frac{16 \sigma_{\rm SB} T^3}{3 \rho^2 \kappa c_p  },
 \end{equation}

\noindent where  $\sigma_{\rm SB}$ is the Stefan–Boltzmann constant, $T$ is internal temperature, $\kappa$ is the effective opacity (which includes thermal conduction), and $c_p$ is specific heat at constant pressure.
  
We assume radiative damping always operates. In contrast, nonlinear wave breaking is assumed to only occur when the wave reaches a large enough radial displacement $\xi_r$ such that $|\xi_r| |k_r| \geq 1$. See Appendix \ref{Wave breaking} for the calculation of $\xi_r$.   We find that wave breaking only occurs in simulations at orbital periods $\lesssim 15 $ min, with radiative damping providing all the dissipation at longer periods. Similarly, we define $m_{\rm break}$ as the characteristic mass scale for breaking, related to characteristic length scale $h_{\rm break}$ by 

\begin{equation}
   m_{\rm break} = 4 \pi \rho r^2 h_{\rm break},
\end{equation}
  
\noindent with $h_{\rm break}$ estimated to be one pressure scale height. Therefore, our full expression for the power deposited per unit mass is

  \begin{equation} \label{change in L}
    \frac{d L_{\rm tide}}{ dm } = -\frac{L_{\rm tide}}{m_{\rm damp}} - \frac{L_{\rm tide}}{m_{\rm break}} \Theta(|\xi_r| |k_r| - 1)
\end{equation}

\noindent where $\Theta$ is simply a Heaviside step function such that wave breaking only contributes when $|\xi_r| |k_r| \ge 1$.

  Figure \ref{fig:internal_wave} shows a radial profile for the heat deposited in the outer region of a WD model at a short orbital period. The outgoing wave is assumed to originate at the bump in $N^2$, so heating only exists exterior to this point. For this snapshot, dissipation is only due to radiative damping. The heating per unit mass reaches a maximum close to the surface of the WD, where $N^2$ is large and the density is low. This behavior is typical for our simulations.

  We remove energy from the wave and deposit it as heat by applying Eq. \ref{change in L}  until either $L_{\rm tide}$ is zero from losses or the end of the gravity wave propagation cavity is reached. In the first case, the energy injected into the WD $L_{\rm heat}$ is equal to $L_{\rm tide}$. In the second case, $L_{\rm heat}/L_{\rm tide}$ will be less than unity, meaning that not all the wave's energy is deposited inside our model. If this fraction is significantly less than unity, the wave carries a significant amount of energy when it reaches the edge of the propagation cavity and likely reflects. In this case, the assumption that the tidal response is in the traveling wave regime breaks down. Instead, standing wave g-modes of discrete frequencies will likely be excited \citep{fuller_tidal_2011, yu_tidally_2021}. 

  \subsubsection{Traveling wave regime: validity} 

\label{wave regime sec}

 \begin{figure}
    \centering
    \includegraphics[scale=0.6]{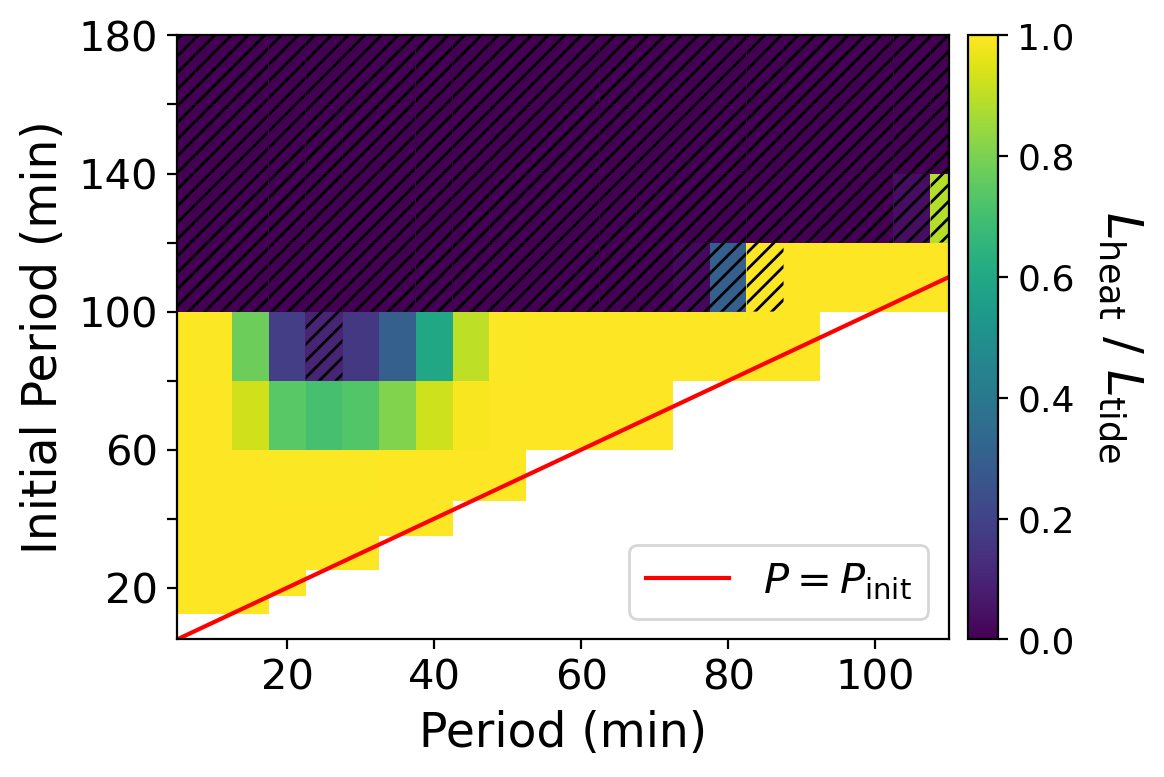}
        \caption{The fraction of the wave energy $L_{\rm tide}$ that is deposited as heat $L_{\rm heat}$ (color shading) within evolving WD models of 0.25 $M_\odot$ (with a $0.3 \, M_\odot$ companion). A given simulation starts on the red line and moves from right to left horizontally, towards shorter orbital periods as inspiral occurs.
        If $L_{\rm heat}/L_{\rm tide}$ is less than unity, the wave is encountering the upper edge of the propagation 
        cavity, where the wave frequency either rises above the Lamb frequency (unhatched region) or the Brunt–Väisälä frequency (hatched region). The traveling wave approximation is good in the yellow regions of this diagram.  }

    \label{fig:wave regime 0.25}
\end{figure}

\begin{figure}
    \centering
    \includegraphics[scale=0.6]{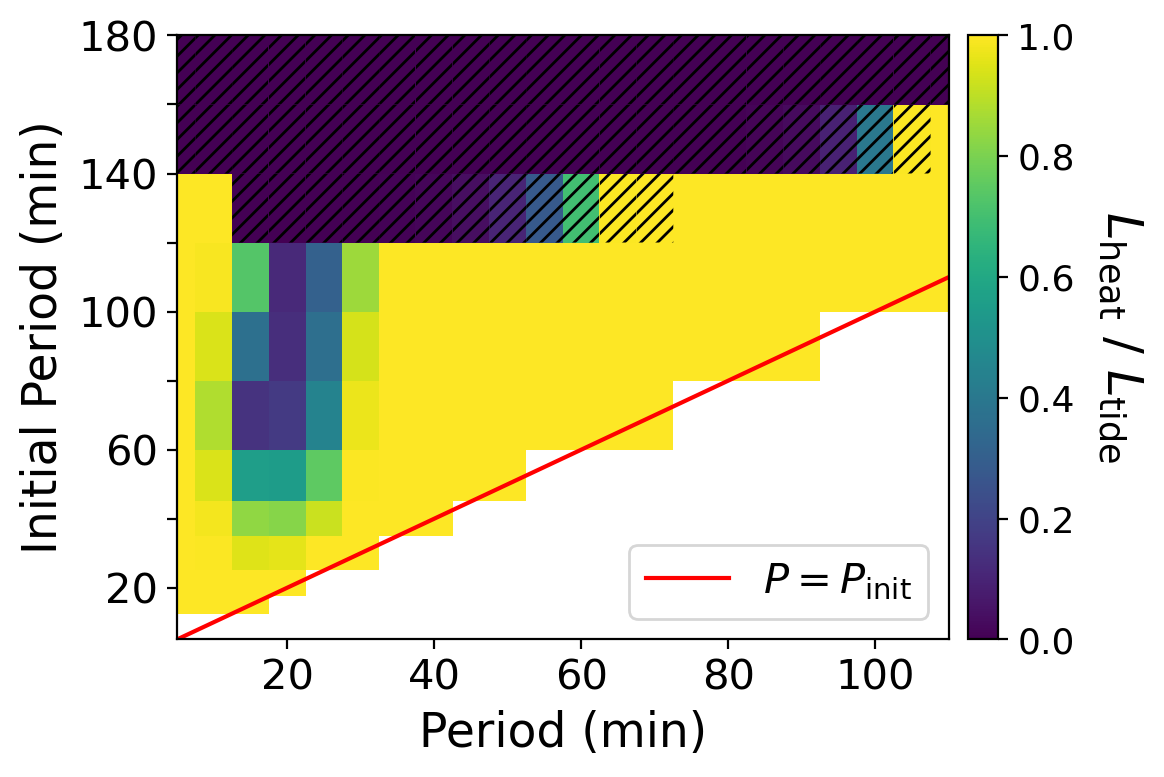}
        \caption{Same as Fig. \ref{fig:wave regime 0.25}, but for a 0.35 $M_\odot$ WD model in a binary. }

    \label{fig:wave regime 0.35}
\end{figure}

\begin{figure}
    \centering
    \includegraphics[scale=0.6]{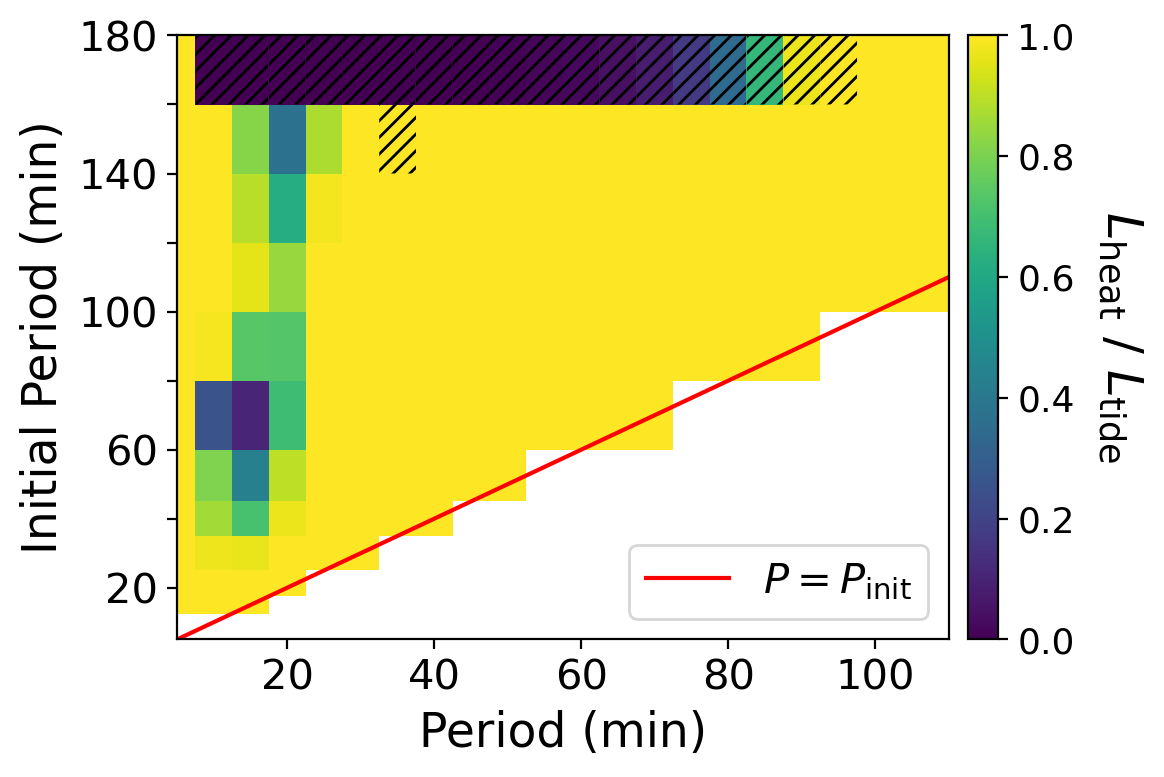}
        \caption{Same as Fig. \ref{fig:wave regime 0.25}, but for a 0.45 $M_\odot$ WD model in a binary.}

    \label{fig:wave regime 0.45}
\end{figure}

  We do not model the standing wave regime, but instead estimate for which models our traveling wave assumption breaks down.  For 3 different WD masses, Figs. \ref{fig:wave regime 0.25}, \ref{fig:wave regime 0.35}, \ref{fig:wave regime 0.45} show a colormap of $L_{\rm heat}/L_{\rm tide}$ versus the parameter space of initial period and orbital period. A given simulation has a fixed initial period, but decreases in period, starting on the red line and moving horizontally right to left. A brighter color corresponds to  $L_{\rm heat}/L_{\rm tide}$ being closer to unity and the traveling wave assumption justified. When $L_{\rm heat}/L_{\rm tide}$ is close to zero, with a darker shading, our model likely breaks down. As a model moves from right to left on the plot, the simulation may encounter regions where the traveling wave regime is more or less appropriate. The most self-consistent simulations occur for horizontal slices that are completely yellow in color, where all the wave power is always deposited. When $L_{\rm heat}/L_{\rm tide} < 1$, the wave is either encountering the edge of the cavity due to $\omega^2$ rising above $L_l^2$ (unhatched, ``Lamb cavity") or $\omega^2$ rising above $N^2$ (hatched, ``Brunt cavity").
  
  There are several trends evident in Figs. \ref{fig:wave regime 0.25}, \ref{fig:wave regime 0.35}, \ref{fig:wave regime 0.45}. The hatched ``Brunt cavity" region occurs for cool WDs in binaries initialized at long orbital periods, because WDs with  $T_{\rm eff} \lesssim 10,000 K$ develop a deep surface convective zone with $N^2<0$. Less massive He-core WDs cool faster and have slower orbital decay, meaning that the hatched region is pushed to lower initial periods for lower mass WDs. 
  
 The darkest unhatched ``Lamb cavity" regions tend to occur for orbital periods of $\approx$15 to 30 min, where the tidal frequency $\omega$ is largest, which moves the cavity's outer boundary inward and also increases the lengthscale for damping (Eq. \ref{damping}). However, at shorter periods, two effects can occur: first, the WD temperature begins to increase due to tidal heating, meaning that radiative damping becomes stronger, which in turn makes more heat dissipate out of the wave before it reaches the cavity's edge. Second, the increasing power carried by the wave means wave breaking is more likely to occur and dissipate more energy within the cavity. These effects are responsible for the yellow region at short orbital periods, where models are nominally again in a traveling wave regime, despite our traveling wave model potentially failing at earlier periods in their evolution.
 At a given period, WDs are cooler for a longer initial orbital period, and radiative damping will be weaker, explaining why, in general, colors are darker moving vertically upward within the unhatched ``Lamb cavity" region. 

  In general, we find that there is still a fairly large parameter space where the traveling wave regime is valid. WDs in binaries with shorter initial orbital periods are more likely to be in this regime.
  Higher mass WDs have a wider range of initial orbital periods where the traveling wave approximation is valid.
  Previous work  has demonstrated that most DWDs are likely born from the common envelope channel at  orbital periods shorter than 2 hours \citep{brown_most_2016},
  and many at less than 1 hour \citep{scherbak_white_2023}. Therefore, the traveling wave regime is likely a good approximation for the evolutionary histories of some systems.  In the remainder of this work, we run models by implementing the traveling wave response and only depositing energy within the propagation cavity, but discuss which models are most valid.

\begin{figure}
    \centering
    \includegraphics[scale=0.6]{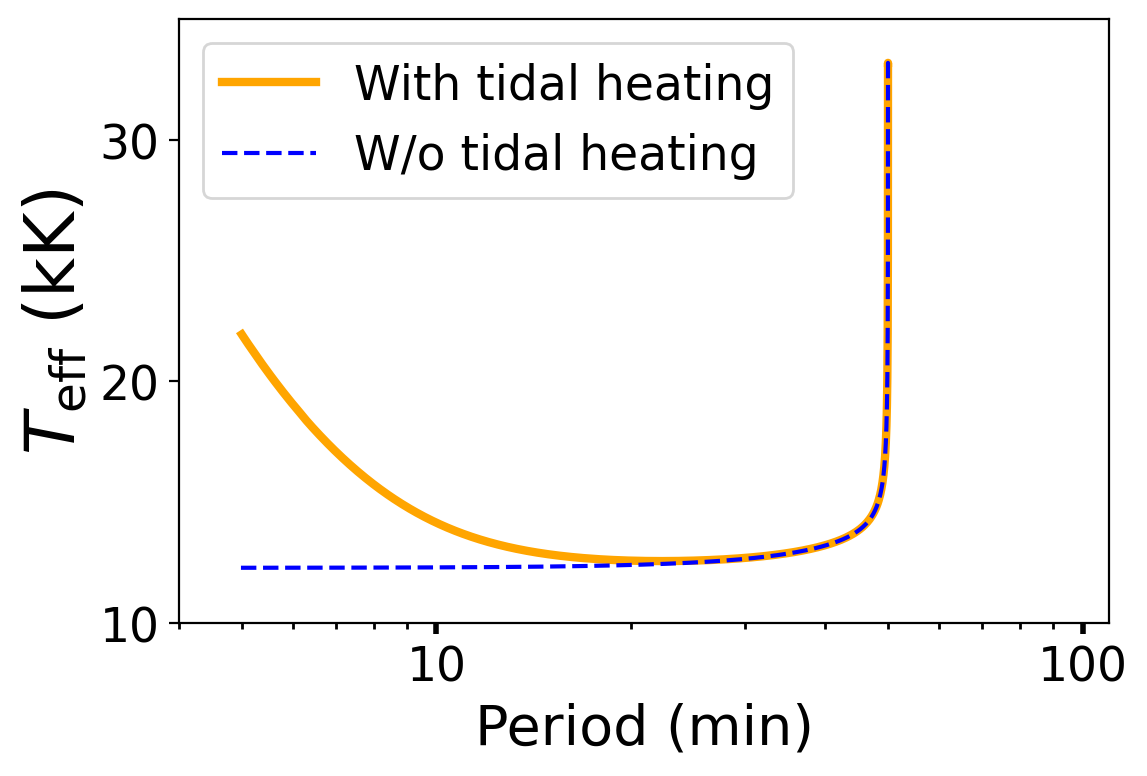}
    \caption{The effective temperature for a 0.25 $M_\odot$ WD, with and without tidal heating, as a function of orbital period. The model moves from right to left towards shorter orbital periods as the binary inspirals. The companion mass is fixed at 0.3  $M_\odot$.
}
    
    \label{fig:Lheat}
\end{figure}


\section{Tidal heating: simulations and observed systems}

\label{Tidal heating compare}

\begin{deluxetable*}{ccccccc}

\tablecaption{Properties of the WD binaries we model. \label{tab: observed systems}}
\tablewidth{0pt}
\tablehead{
\colhead{Name} & \colhead{$P_{\rm orb}$ (min)} & \colhead{$M_1$ $(M_\odot)$} &
 \colhead{$M_2 (M_\odot)$} & \colhead{$T_{\rm eff,1 }$ (K)} & \colhead{$T_{\rm eff,2 }$ (K)} & \colhead{$P_{\rm birth}$ (min)}
}
\decimalcolnumbers
\startdata
ZTF J1539+5027$^{\textcolor{red}{1}}$ & 6.91 & $0.21^{+0.014}_{-0.015}$ & $0.61^{+0.017}_{-0.022}$ &$<10,000$  & $48,900 \pm 900$ & not modeled     \\
 ZTF J2243+5242$^{\textcolor{red}{2}}$ & 8.80 & $ 0.323^{+0.065}_{-0.047}   $ (LC)   & $0.335^{+0.052}_{-0.054}$ (LC)    & $26,300^{+1700}_{-900}$ (SED)    & $19,200^{+1500}_{-900}$ (SED)  & 10 - 26   \\
  &  & $0.317^{+0.074}_{-0.074}$ (Spect)   & $0.274^{+0.047}_{-0.047}$ (Spect)    & $26,520^{+130}_{-130}$ (Spect)    & $19,670^{+100}_{-100}$ (Spect)  &   \\
SDSS J0651+2844$^{\textcolor{red}{3,4,7}}$& 12.75 & $0.247^{+0.015}_{-0.015}$ & $0.49^{+0.02}_{-0.02}$ & $16,530  \pm 200$   & $8,700  \pm 500$ & 20 - 30    \\
ZTF J0538+1953$^{\textcolor{red}{5}}$ & 14.44 & $0.45^{+0.05}_{-0.05}$ & $0.32^{+0.03}_{-0.03}$  & $26,450 \pm 725$ & $12,800 \pm 200$ & 36 - 44  \\
SDSS J0935+4411$^{\textcolor{red}{6,7}}$ & $19.80$ & $0.312^{+0.019}_{-0.019}$ & $0.75^{+0.24}_{-0.23}$  & $21,660 \pm 380$ & n/a  & $22 - 41$  \\
\enddata
\tablecomments{Parameters for all the systems we model, including the orbital period, $P_{\rm{orb}}$, the masses of the two components, \(M_1\) and \(M_2\), and the effective surface temperature of the components, $T_{\rm{eff,1}}$ and $T_{\rm{eff,2}}$.  Measured parameters are from  \protect \cite{burdge_general_2019}$^{\textcolor{red}{1}}$, 
  \protect \cite{burdge_88_2020}$^{\textcolor{red}{2}}$, \protect \cite{brown_12_2011}$^{\textcolor{red}{3}}$, \protect \cite{hermes_rapid_2012}$^{\textcolor{red}{4}}$,  
  \protect \cite{burdge_systematic_2020}$^{\textcolor{red}{5}}$, \protect \cite{kilic_new_2014}$^{\textcolor{red}{6}}$, \protect \cite{brown_most_2016}$^{\textcolor{red}{7}}$. The last column is our estimate of the birth period, the post-common envelope orbital period when the younger WD formed (see text for the analysis). }

\end{deluxetable*}

\begin{figure*}
    \centering
    \includegraphics[scale=.55]{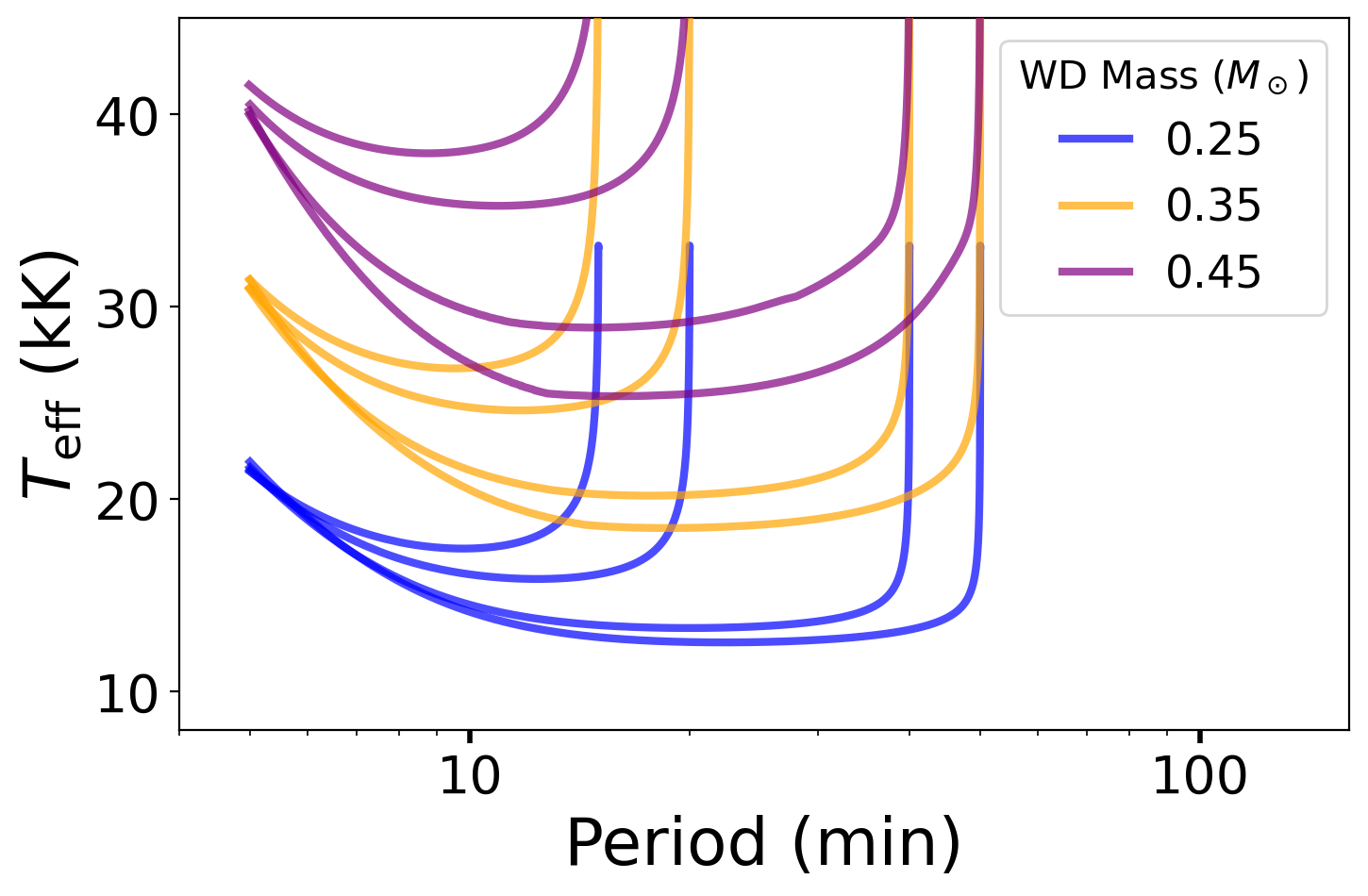}
    \caption{ The evolution of effective temperature versus orbital period for a grid of WD models undergoing tidal heating. Models of same color are of the same mass. The models begin evolution in a binary at orbital periods of 15, 20, 40, and 50 min. The companion mass is fixed at 0.3 $M_\odot$.  }
    \label{fig:Teff_grid}
\end{figure*}

Fig. \ref{fig:Lheat} demonstrates the effect of tidal heating on a WD's surface temperature as the DWD binary evolves from 50 min to shorter orbital periods. Note from Fig. \ref{fig:wave regime 0.25} that this simulation (WD mass 0.25 $M_\odot$, initial period 50 min) is likely always in the traveling wave regime. The magnitude of the heating increases with decreasing orbital period because the tidal potential driving the wave is stronger. The temperature of the primary WD is predicted to be much higher at very short periods (a difference of $>$10,000 K at an orbital period of 5 min) compared to the case where tidal heating is neglected. We find that there is a negligible increase in the radii of our tidally heated models. Significant heating begins at periods $\lesssim$ 15 minutes, corresponding to a gravitational wave inspiral time $\lesssim$ 2 Myr for the component masses. Thus, the heating generally occurs over a short timescale compared to the WD's age. In the following, we present similar results for a grid of WD models.

Fig. \ref{fig:Teff_grid} shows tidal heating for a grid of WDs of different masses initialized in binaries of different initial orbital periods. For simplicity, we show initial orbital periods of 15 to 50 minutes, although our full grid ranges from 15 to 200 minutes.
For the 0.35 and 0.45 $M_\odot$ models initialized at 50 minutes, the traveling wave regime starts to break down (this is the source of the kink for the bottom-most 0.45 $M_\odot$ curve at about 13 minutes, where, from Fig. \ref{fig:wave regime 0.45}, there is a change from a small fraction of $L_{\rm tide}$ being deposited to a larger fraction being deposited). However, the traveling wave regime is likely a good approximation for the other models in this plot.

The power carried by the wave, ${L}_{\rm tide}$, scales with WD mass approximately as $M_1^{17/6}$ \citep{fuller_dynamical_2013}, so more massive WDs experience greater heating at short periods and have higher temperatures, regardless of their initial orbital period. In addition, more massive WDs have smaller radii and therefore the effect of tidal heating on $T_{\rm eff}$ will be more pronounced.

\begin{figure}
    \centering
    \includegraphics[scale=.45]{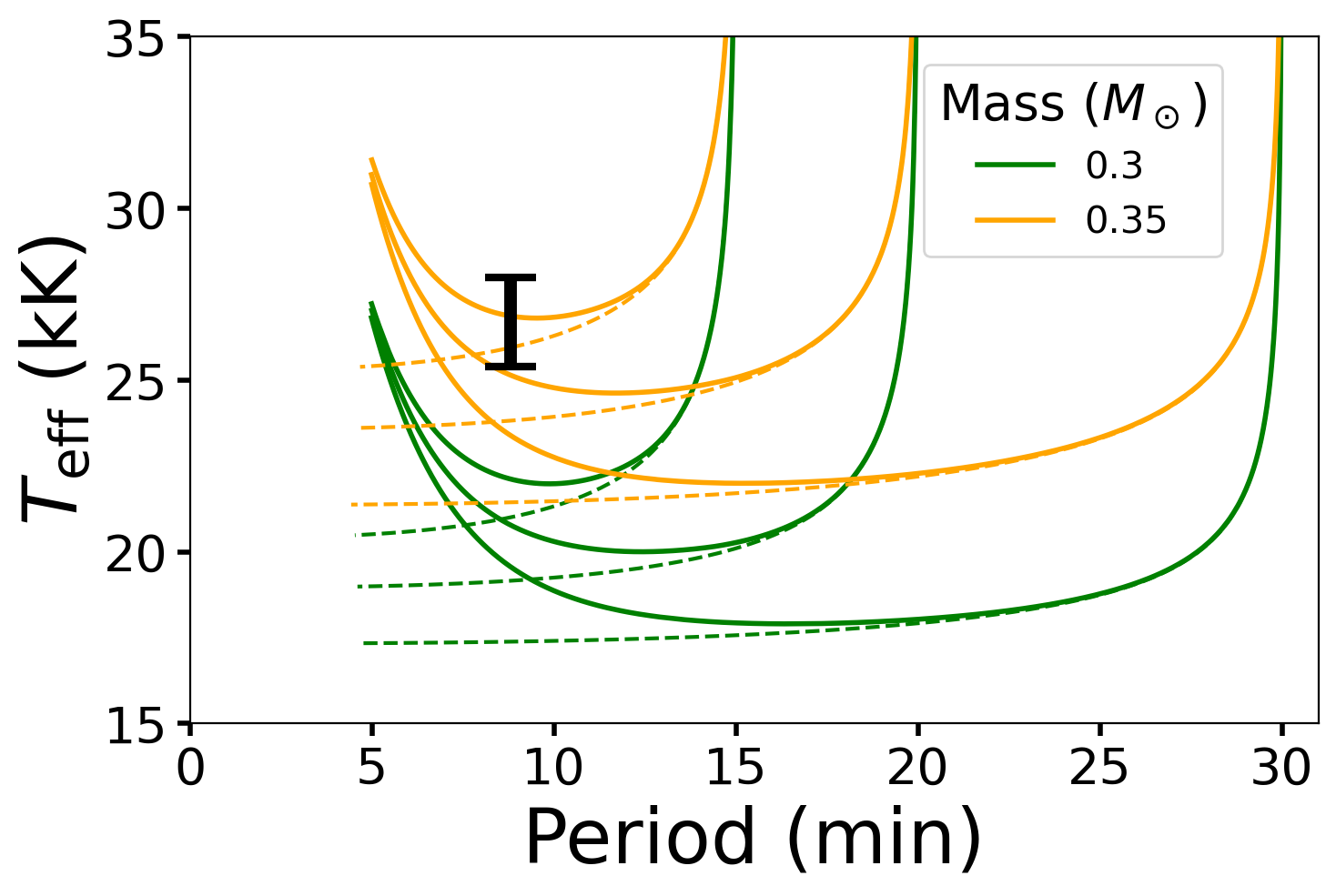}
    \caption{ Effective temperature versus orbital period for a grid of WD simulations, compared to the observed temperature (black errorbar) of ZTF J2243's hotter component (mass $\approx 0.32 M_\odot$). The models vary in WD mass (differing colors) and initial orbital period of the system, whereas companion mass is fixed at $0.3 M_\odot$. Dashed lines are models with the same parameters, but have tidal heating turned off. }
    \label{fig:Teff_compare_1}
\end{figure}


\begin{figure}
    \centering
    \includegraphics[scale=.45]{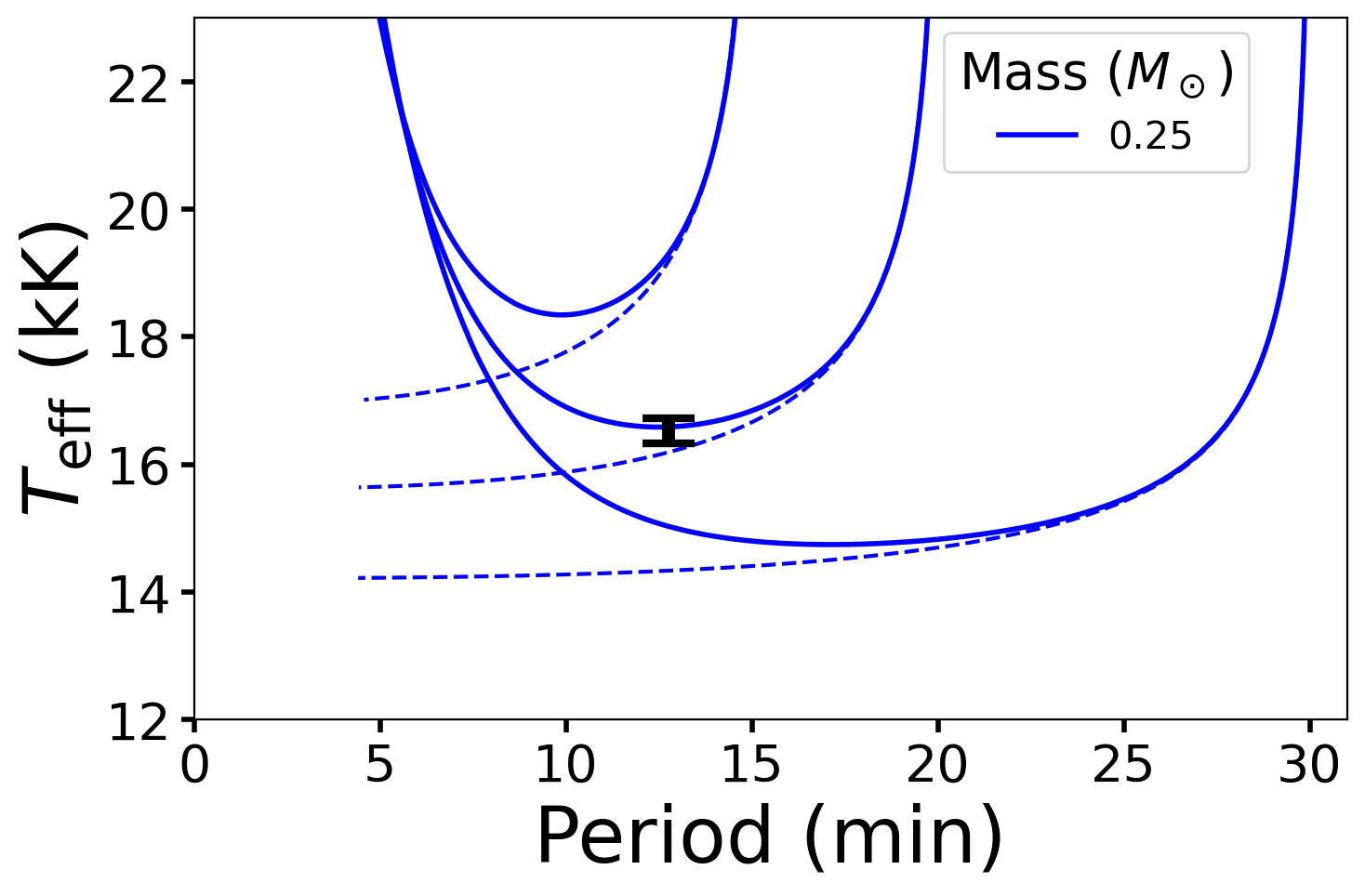}
    \caption{Similar to Fig. \ref{fig:Teff_compare_1}, but for SDSS J0651.  The mass of the hotter component, which we fit to, is $\approx 0.25 M_\odot$. The companion mass is fixed at $0.5 M_\odot$.  }
    \label{fig:Teff_compare_2}
\end{figure}

\begin{figure}
    \centering
    \includegraphics[scale=.45]{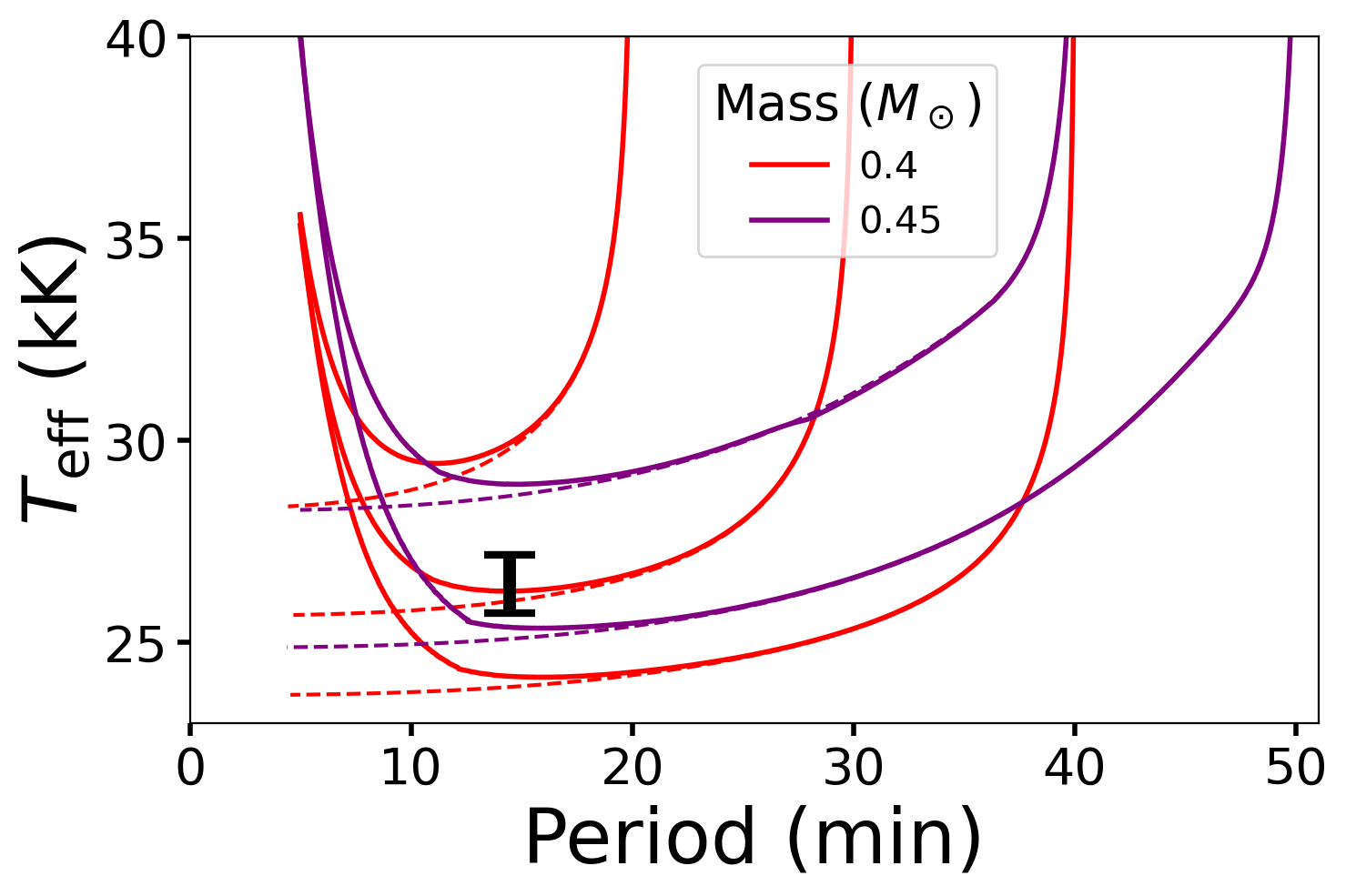}
    \caption{Similar to Fig. \ref{fig:Teff_compare_1}, but for ZTF J0538. The mass of the hotter component, which we fit to, is $\approx 0.45 M_\odot$. The companion mass is fixed at $0.3 M_\odot$.  }
    \label{fig:Teff_compare_3}
\end{figure}

\begin{figure}
    \centering
    \includegraphics[scale=.45]{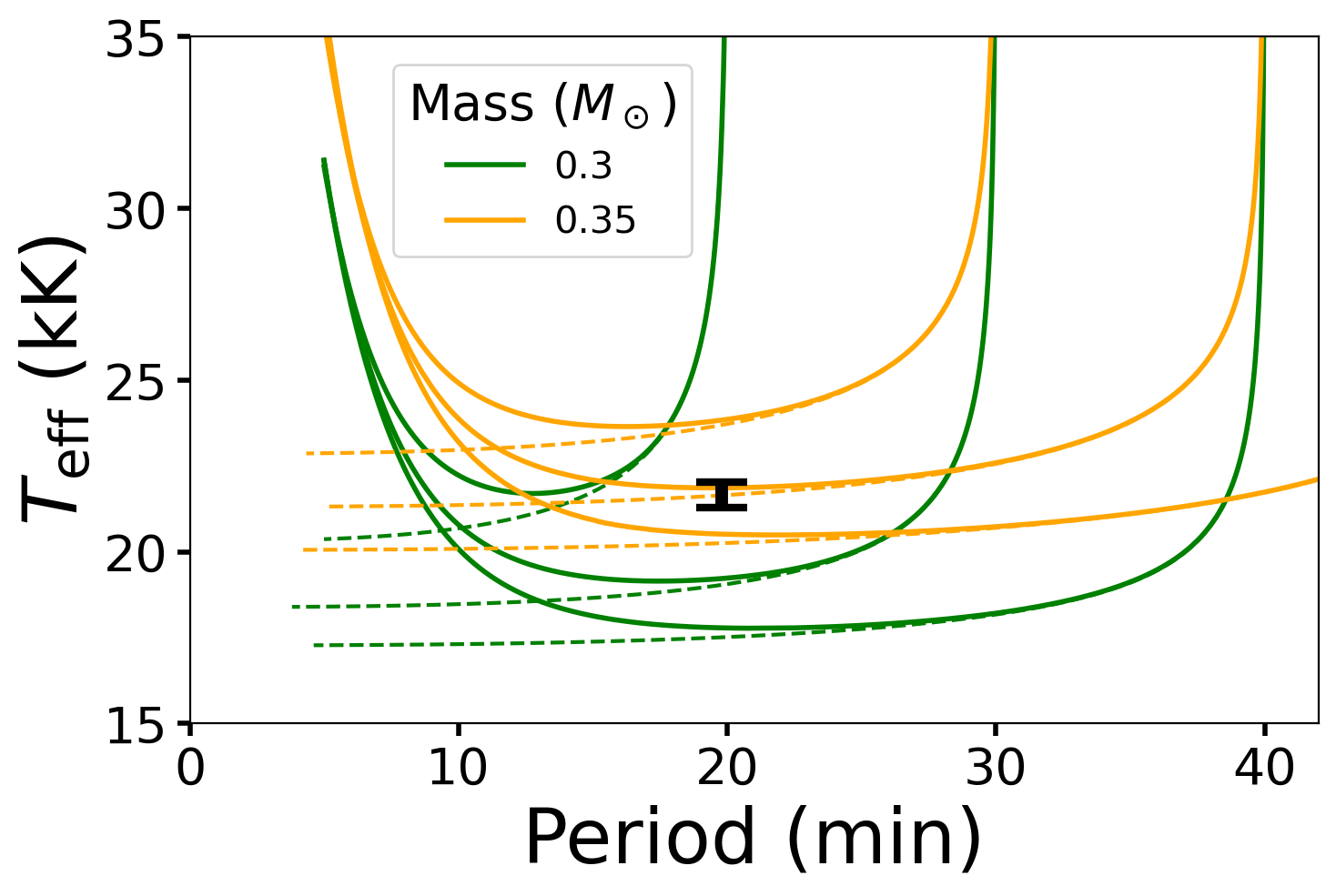}
    \caption{Similar to Fig. \ref{fig:Teff_compare_1}, but for SDSS J0935.  The mass of the hotter component, which we fit to, is $\approx 0.31 M_\odot$. The companion mass is fixed at $0.75 M_\odot$.  }
    \label{fig:Teff_compare_4}
\end{figure}

We also compare our quantitative predictions for $T_{\rm eff}$ to observed ultrashort-period double white dwarf systems. The properties of these systems are summarized in Table \ref{tab: observed systems}. Note also that these are the 5 detached shortest-period LISA verification binaries systems discussed in \cite{kupfer_lisa_2023}. In this work, we only simulate tidal heating in the younger, more recently-formed WD, treating the other as a point mass. If only one WD has a mass consistent with a He-core WD, we assume that is the younger because its progenitor star likely evolved second (although this may not always be true). If there are two He-core WDs, we assume the hotter one is the younger. In Table \ref{tab: observed systems}, we define $M_1$ as the WD we assume is younger and for which we are modeling tidal heating.

Figs. \ref{fig:Teff_compare_1}
- \ref{fig:Teff_compare_4} show a comparison of our WD models, with and without tidal heating turned on, to the observed $T_{\rm eff}$ of systems in Table \ref{tab: observed systems}. We do not seek to exactly find the best set of models that match to the observed characteristics, but compare relevant models with and without tidal heating and quantify the difference between them.

In Fig. \ref{fig:Teff_compare_1}, showing the 8.8 minute system ZTF J2243, the models that best match the observed $T_{\rm eff}$ of the WD primary  correspond to $0.35$ $M_\odot$ WDs. Both simulations with and without tidal heating can match the data - for instance, the two $0.35$ $M_\odot$ models starting at a 15 min period are consistent with the error bar. Models starting at 30 min or greater are not hot enough to match the data, even with tidal heating turned on. Therefore, a WD that cools sufficiently is unable to be heated by tides to the observed hot temperature ($\approx$ 26,000 K). Comparing the hottest 2 pairs of 0.35 $M_\odot$ models (the top 2 pairs of orange curves), the increase in luminosity is about 20\% with tidal heating implemented. Thus, tidal heating is a relatively small fraction of the WD primary's energy output.

Figs. \ref{fig:Teff_compare_2} and \ref{fig:Teff_compare_3} show similar results for the 12.8 minute system SDSS J0651 and 14.4 minute system ZTF J0538. In both cases, the difference between pairs of models with and without tidal heating at the observed orbital period is relatively small, with the change in $T_{\rm eff}$ $<$1,000 K and the change in luminosity between 5-10\%.
For J0651, models starting at 20 min match well, but models starting at 15 minutes are too hot and models starting at 30 minutes are too cool. For J0538, models starting at 30 minutes match well, but models starting at 20 minutes are too hot and models starting at 40 minutes are too cool. Accounting for tides, therefore, does not appear to greatly alter the parameter space of initial periods that can replicate these systems. 

Finally, Fig. \ref{fig:Teff_compare_4} shows the 19.8 minute system SDSS J0935, the longest period binary we compare to, and demonstrates that there is little difference between tidally heated and non-tidally heated models at orbital periods near 20 min.

Note that the models in Figs. \ref{fig:Teff_compare_1}-\ref{fig:Teff_compare_4} have short initial periods and are therefore mostly in the traveling wave regime. An exception is the higher-mass, longer initial period (50 min) models in Fig. \ref{fig:Teff_compare_3}, for which our treatment may start to break down. Models born at longer orbital periods than plotted are likely in the standing wave regime, and are unlikely to undergo significantly more heating compared to those in the traveling wave regime. Therefore, since some traveling wave models already fall below the observed $T_{\rm eff}$, the standing wave models are unlikely to further the space of models that can match the observed $T_{\rm eff}$ of these four hot WD primaries. 

We do not show a plot for the 7-minute binary ZTF J1539, for which the He-WD component has a $T_{\rm eff}$ of $<$10,000 K. The history of this system is uncertain, with mass transfer from the He-WD to its companion potentially playing a role \citep{burdge_general_2019}. It may have also formed via the stable mass transfer channel rather than the CE channel \citep{li_2019,chen_2022}. Our models tend to predict significantly higher $T_{\rm eff}$ than 10,000 K, regardless of initial orbital period, meaning that our traveling-wave models are incompatible with the low observed temperature. However, we note that if a WD were to cool to $<$10,000 K the tidal response is likely in the standing wave regime, which we do not attempt to model but will involve resonances with g-modes. In our traveling wave models, the tidal power is deposited near the surface of the WD. In the standing wave regime, the tidal heating power could be smaller, and the location of heat deposition could potentially be deeper inside the WD. This could allow for much cooler WDs.

\section{Discussion}

\label{discussion}

\subsection{Implications for common envelope efficiency}

Our analysis shows that, for the systems in Table \ref{tab: observed systems}, tides are not transforming a cool WD into a hot one, but instead are only slightly heating already hot WDs. Tidal heating does not significantly alter estimates of the initial orbital period of these ultrashort period DWD binaries. To better understand the formation history of these systems, it is reasonable to use a grid of non-tidally heated models, with varying mass and H envelope mass, to estimate the post-common envelope (CE) periods of these binaries. We performed similar analysis for several DWD systems in  \cite{scherbak_white_2023}. We follow the same procedure and only briefly outline it here.

We fit a grid of \texttt{MESA} models to the observed radii and $T_{\rm eff}$ of the WD primary to estimate the birth period. As this is also the post-CE period, we thereby obtain the post-CE orbital energy $E_{ \rm{orb, f}}$. We estimate the pre-CE orbital energy  $E_ {\rm{orb, i}}$ and the envelope binding energy $E_{\rm bind}$ using a grid of red giant progenitor models whose core mass matches the observed He-WD mass. We then combine these to estimate the common envelope efficiency $\alpha_{\rm CE}$, defined by

\begin{equation}
    \alpha_{\rm CE} = \frac{E_{\rm bind}}{E_{ \rm{ orb, f}} - {E_ {\rm orb, i}} }.
\end{equation}

Our estimates of the birth period are shown in the rightmost column of Table \ref{tab: observed systems}. The birth periods are quite short, significantly less than 1 hour, in agreement with the conclusion of \cite{brown_most_2016} that most low-mass DWD binaries are born at short orbital periods. It also means that the DWD binaries we model are young.

\begin{figure*}
    \centering
    \includegraphics[scale=.5]{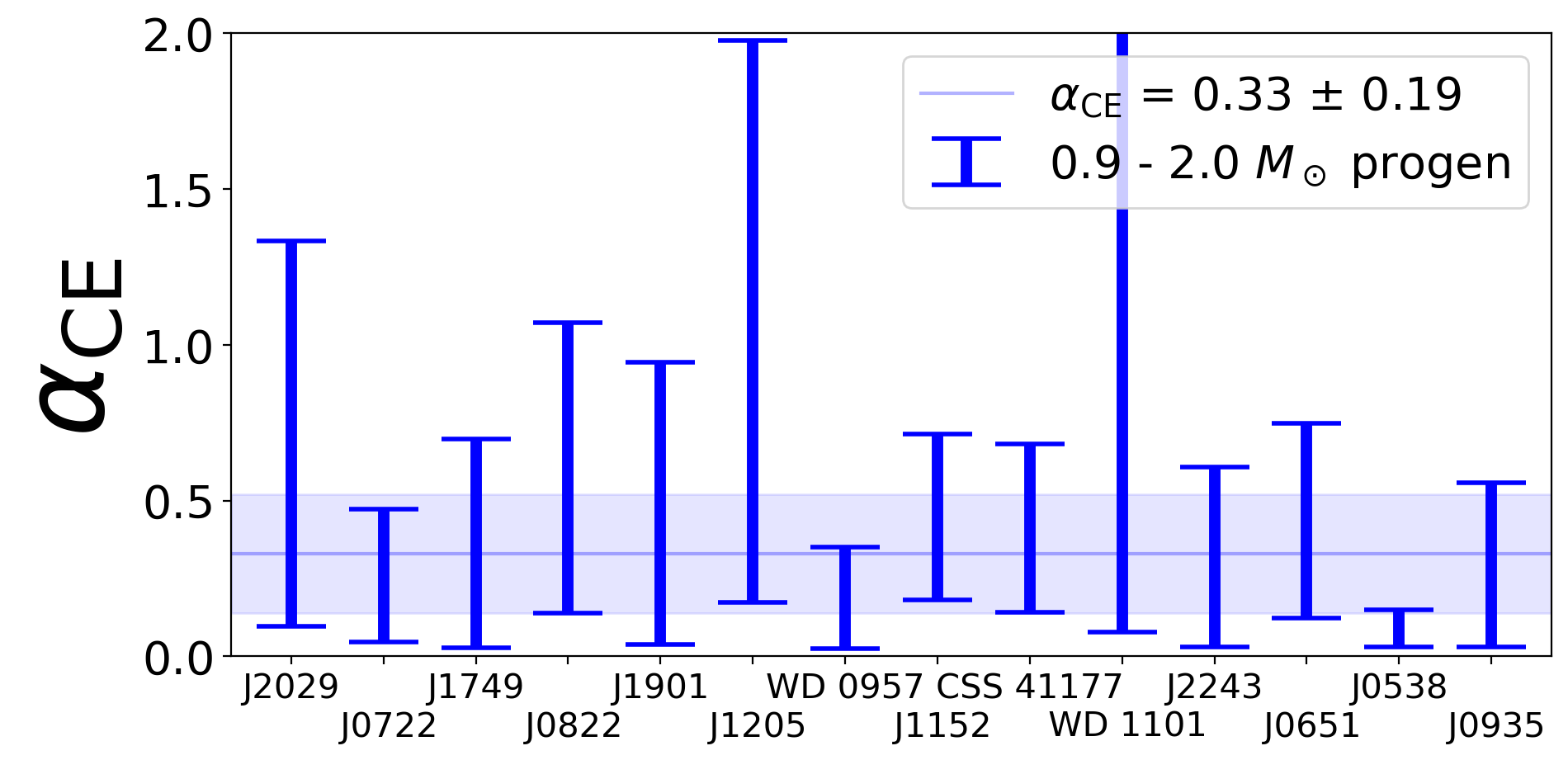}
    \caption{Values of $\alpha_{\rm CE}$ using a grid of He WDs to model the WD primary and assuming the progenitor is a red giant star from 0.9 - 2.0 $M_\odot$. The     
    horizontal line and shaded region denotes a least squares fit to the ensemble of systems, excluding the  
    rightmost 4 (the binaries discussed in this work). The other 10 systems 
    used for the fit are longer period WD binaries modeled in \protect \cite{scherbak_white_2023}. The inferred values of $\alpha_{\rm CE}$ for the short-period binaries in this work are consistent with expectations, apart from J0538 (see text for explanation).}
    \label{fig:alpha}
\end{figure*}

Fig. \ref{fig:alpha} shows the range of $\alpha_{\rm CE}$ for 10 WD binaries modeled in \cite{scherbak_white_2023} and the 4 binaries discussed in Figs.  \ref{fig:Teff_compare_1}-\ref{fig:Teff_compare_4}. The previously-modeled WD binaries are at longer orbital periods $> 20$ min. The best-fit range for $\alpha_{\rm CE}$ using these previous 10 systems is from about 0.15 to 0.5 and overlaps well with 3 of the 4 new systems. For the most part, we find that these ultrashort period binaries, which we previously did not model because of the uncertainty of the role of tidal heating, are consistent with a low common envelope efficiency.

An exception is ZTF J0538, which has an even lower associated range of $\alpha_{\rm CE}$ due to its short post-CE orbital period and higher mass WD, which means the progenitor was more evolved up the red giant branch with a lower $E_{\rm bind}$. However, the mass of the WD primary  ($0.45^{+0.05}_{-0.05} M_\odot$) is quite large for a He WD, and it could instead correspond to a low-mass CO-core WD. This requires a significantly different evolutionary history, likely involving a He-burning sdB phase following the CE \citep{han_2003, xiong_subdwarf_2017}. This would entail a much longer evolution time following the CE event, and hence a longer post-CE orbital period than that listed in Table \ref{tab: observed systems}, and a larger value of $\alpha_{\rm CE}$ than that shown in Figure \ref{fig:alpha}.

\subsection{Application to CO WDs}

This work has focused in tidal heating in He-core WDs, where the companion is either a He or CO WD. A discovery of a short-period binary where both WDs have CO cores would motivate modeling of tidal heating in a CO WD as well. The excitation of outgoing gravity waves in a CO WD \citep{fuller_dynamical_2012} is similar to the He WD case \citep{fuller_dynamical_2013}, although the primary location of the gravity wave excitation is instead at the carbon-helium composition gradient. Therefore, most of our modeling can be directly extended to \texttt{MESA} simulations of CO WDs.

The most significant difference would likely be the parametrization of the tidal torque through $F(\omega)$ (Eqs. \ref{torque} and \ref{f_omega}), which we could determine by running the numerical waveform solver on CO WD models. In this work, we found smaller values of $F(\omega)$ compared to those in \cite{fuller_dynamical_2013}, which used older He WD models. If this trend holds true for CO WDs as well, the results of \cite{fuller_dynamical_2012} will be changed somewhat. In particular, the critical period when spin synchronization begins may be shorter, and the magnitude of tidal heating may be higher, than presented in \cite{fuller_dynamical_2012}. 
Overall, we expect similar qualitative behavior for CO WD models, with tidal heating becoming significant at periods $\lesssim$ 10-20 min and becoming the dominant source of luminosity pre-merger.


\subsection{Potential future observations}

The discovery of more ultrashort-period double WD binaries would be invaluable to add to our comparison sample. The enhancement in temperature due to tidal heating increases for shorter orbital periods, so systems with orbital periods $<$ 10 min are better for distinguishing tidally heated and non-tidally heated models. Specifically, values of $T_{\rm eff}$ tend to converge for a given WD mass (Fig. \ref{fig:Teff_grid}) at the shortest orbital periods pre-mass transfer, which would be a testable prediction. In addition, discovering more systems at similar periods to the 7-minute binary ZTF J1539 would shed light on whether ZTF J1539's cool component is an outlier, perhaps due to its low mass, or if there are additional cool WDs for which the tidal heating may be more complicated than modeled here. 

In this work, we have only compared to measured values of $T_{\rm eff}$. Measured spin values would be another way to test our models, which predict spin synchronization as a function of orbital period. In some cases, the projected rotation speed, $v \sin i$, could potentially be measured. For complete synchronization, we expect
\begin{equation}
    v \sin i \simeq 200 \, {\rm km}\,{\rm s}^{-1} \bigg(\frac{R_{\rm WD}}{2 \times 10^9 \, {\rm cm}} \bigg) \bigg(\frac{P_{\rm orb}}{10 \, {\rm min}}\bigg)^{\!-1}
\end{equation}
for an edge-on system. This could possibly be measured from high-resolution spectra resolving the non-LTE line cores of DA WDs, or helium lines of DB or DAB WDs.

\section{Conclusion}

In this paper, we have modeled tidal heating in double WD binary \texttt{MESA} simulations, where the dynamical tidal excitation occurs within helium-core WDs and the companion is a point mass. We approximate the tidal response as traveling gravity waves that are damped in the outer layers of the WD. We run a grid of models varying \textbf{1}: the mass of the He-core WD $M_1$ \textbf{2}: the mass of the companion $M_2$ and \textbf{3}: the post-common envelope orbital period of the WD binary when the second WD formed.

In our simulations, the binary inspirals to shorter periods due to the emission of gravitational waves. At every time step, we calculate the angular momentum and energy flux carried by the tidally excited waves, which leads to spin-up and heating of the WD. For our grid of WD models, we calibrate the strength of the tidal response using the numerical solver of \cite{fuller_dynamical_2013}. 

The WDs' spins become increasingly synchronized with the orbit at orbital periods $\lesssim$ 30 minutes. We model heat deposition in the outer layers of the WD, accounting for damping due to radiative diffusion and nonlinear wave breaking. We self-consistently compute the changing WD structure, heating rate, and heating location. The inner edge of wave propagation is the transition between the He-core and H envelope, and the outer edge is where the wave becomes evanescent. We determine the fraction of the wave's energy which has been deposited before it reaches the cavity's outer edge. If the fraction is low, our traveling wave approximation breaks down and the response will instead likely be a standing wave. Figs. \ref{fig:wave regime 0.25} - \ref{fig:wave regime 0.45} demonstrate where the traveling wave response is appropriate. In general, hot and young WDs, corresponding to binaries with short initial orbital periods $\lesssim$ 2 hour, are more likely to be in the traveling wave regime. The traveling wave regime extends to longer initial periods for higher mass WDs ($\sim$ 0.45 $M_\odot$) compared to lower mass WDs ($\sim$ 0.25 $M_\odot$).

At orbital periods $\gtrsim$ 20 min, the WDs follow a cooling track and tidal heating is negligible. However, at shorter orbital periods, the energy flux carried by the tidally excited gravity waves becomes large enough to cause substantial heating. Compared with a non-tidally heated model, the enhancement in $T_{\rm eff}$  can exceed 10,000 K at periods of 5-10 min. When tidal heating dominates, WDs of the same mass tend to converge to similar values of $T_{\rm eff}$, despite differences in their previous cooling history. 

We compare our tidal heating models to the five shortest-period double WD systems known, ranging from 8 to 20 min orbital periods.
Our tidally heated models that match well to the measurements are likely in the traveling wave regime, but they only cause a mild temperature increase. Both tidally heated and non-tidally heated models are capable of reproducing the measured surface temperatures, but only when starting at a similar range of orbital periods. Therefore, uncertainty in tidal heating does \textit{not} lead to a large uncertainty in  the binary's initial period. Since tidal heating is a small fraction of these WDs' luminosities, they are intrinsically hot and young, and they were born at periods $\lesssim 45 \, {\rm min}$. The exception to these findings is ZTF J1539's primary, which is cooler than our tidal heating models would predict, but has caveats involving mass transfer, its formation history, and the role of standing waves at its cool temperature.

Finally, we reconstruct the values of the common envelope (CE) efficiency parameter $\alpha_{\rm CE}$ that can replicate these systems. Since tidal heating does not significantly impact estimates of the initial (post-CE) period of these ultrashort-period binaries, we perform the same analysis as \cite{scherbak_white_2023}, which modeled longer period WD binaries.  For reasonable assumptions of the progenitor star mass, we find a fairly low value of $\alpha_{\rm CE}$, in the range 0.1 to 0.5, is necessary to form the ultrashort-period systems. These low values of $\alpha_{\rm CE}$ are consistent with those found in \cite{scherbak_white_2023}.

\label{conclusion}



\begin{acknowledgments}
We are grateful for support from the NSF through grant AST-2205974. JF is thankful for support through an Innovator Grant from The Rose Hills Foundation. We thank the anonymous referee for their useful
feedback and suggestions.
\end{acknowledgments}

\software{MESA \citep{Paxton2011, Paxton2013, Paxton2015, Paxton2018, Paxton2019}, Python.\texttt{MESA} models and files are available in a Zenodo repository \citep{scherbak_2024_10452487}.}

\appendix

\section{AM transport}
\label{AM transport}

This work has assumed that the WD is rigidly rotating. This requires angular momentum (AM) transport between the location where AM is deposited by the outgoing gravity waves, and the rest of the WD's interior. We have not performed detailed modeling of AM transport in \texttt{MESA}. Rather, we have estimated the coupling timescale for AM to be transported by a magnetic dynamo, and compared that to the spin-up timescale of the WD's interior. For coupling from the WD surface at radius $R$ to a location with internal radius $r$, we define the AM coupling timescale $t_{\rm coup}(r)$ as 

\begin{equation} 
     t_{\rm coup}(r) = \int_r^R dr' \frac{(R-r')}{\nu(r')}
\end{equation}

\noindent where $\nu$ is the AM diffusivity found in \cite{fuller_slowing_2019}
 which details a modification to the Tayler-Spruit dynamo \citep{spruit_dynamo_2002}. It is given by

 \begin{equation}
     \nu = \alpha^3 r^2 \Omega_s \left(\frac{\Omega_s}{N_{\rm eff}}\right)^2.
 \end{equation}
 
 \noindent We use $\alpha \approx$ 0.25 \citep{fuller_spins_2022}. In calculating $\nu$, we assume  that $N_{\rm eff}$ is the full Brunt–Väisälä frequency, as that provides a lower limit on $\nu$ and an upper limit on $t_{\rm coup}(r)$.  Because our assumption of rigid rotation breaks down for relatively long coupling times, this makes our calculation more conservative. 

We define the spin-up timescale $ t_{\rm spinup}(r)$ as the timescale to synchronously spin up the layers exterior to $r$ with moment of inertia $I(>r)$. With $\dot{J_z}$ as the full tidal torque as calculated in Eq. \ref{torque},

\begin{equation} 
     t_{\rm spinup}(r) = \frac{\Omega I(>r) }{\dot{J_z}}.
\end{equation}

We have found that the coupling timescale is generally less than the spin-up timescale, supporting the assumption of rigid rotation. This assumption is better for lower mass He-core WDs. The assumption of rigid rotation can break down at longer orbital periods $\gtrsim 30$ min, mainly because the spin rate is smaller and hence the AM diffusivity $\nu$ is smaller. However, at longer orbital periods, tidal effects are weaker and deviation from rigid rotation will likely not matter, especially for tidal heating. We find at short orbital periods the coupling timescale to be shorter than the spin-up timescale at most internal radii $r$, and therefore the WD tends to be rigidly rotating when it is rotating closest to synchronization and undergoing the strongest tidal heating.

\section{Calculation of wave breaking}

\label{Wave breaking}

We calculate the radial displacement $\xi_r$ associated with outgoing gravity waves in order to determine when nonlinear wave breaking is likely to occur. Because the power in the wave $L_{\rm tide}$ decreases as the wave moves outwards (Eq. \ref{change in L}), $L_{\rm tide}$ is a function of internal radius $r$. $L_{\rm tide}(r)$ can be related to $\xi_{\rm \perp}$, the perpendicular displacement, with

\begin{equation}
    L_{\rm tide}(r) =2 \sqrt{l(l+1)}  \frac{\omega^4 \rho  r^3   |\xi_{\perp}|^2  }{ N} 
\end{equation}

\noindent and $\xi_{\rm \perp}$ and $\xi_r$ are related in the WKB approximation via

\begin{equation}
    |\xi_r| / |\xi_{\perp}|  = \sqrt{l(l+1)} \frac{\omega}{N}.   
\end{equation}

\noindent These formulae will break down if the wave nears the edge of the propagation cavity where $\omega^2$ becomes comparable to $L_l^2$ (i.e. if the traveling wave approximation breaks down, Sec. \ref{wave regime sec}).

\section{Effect of hydrogen envelope mass on tidal heating}

\label{H effect sec}

\begin{figure}
    \centering
    \includegraphics[scale=.5]{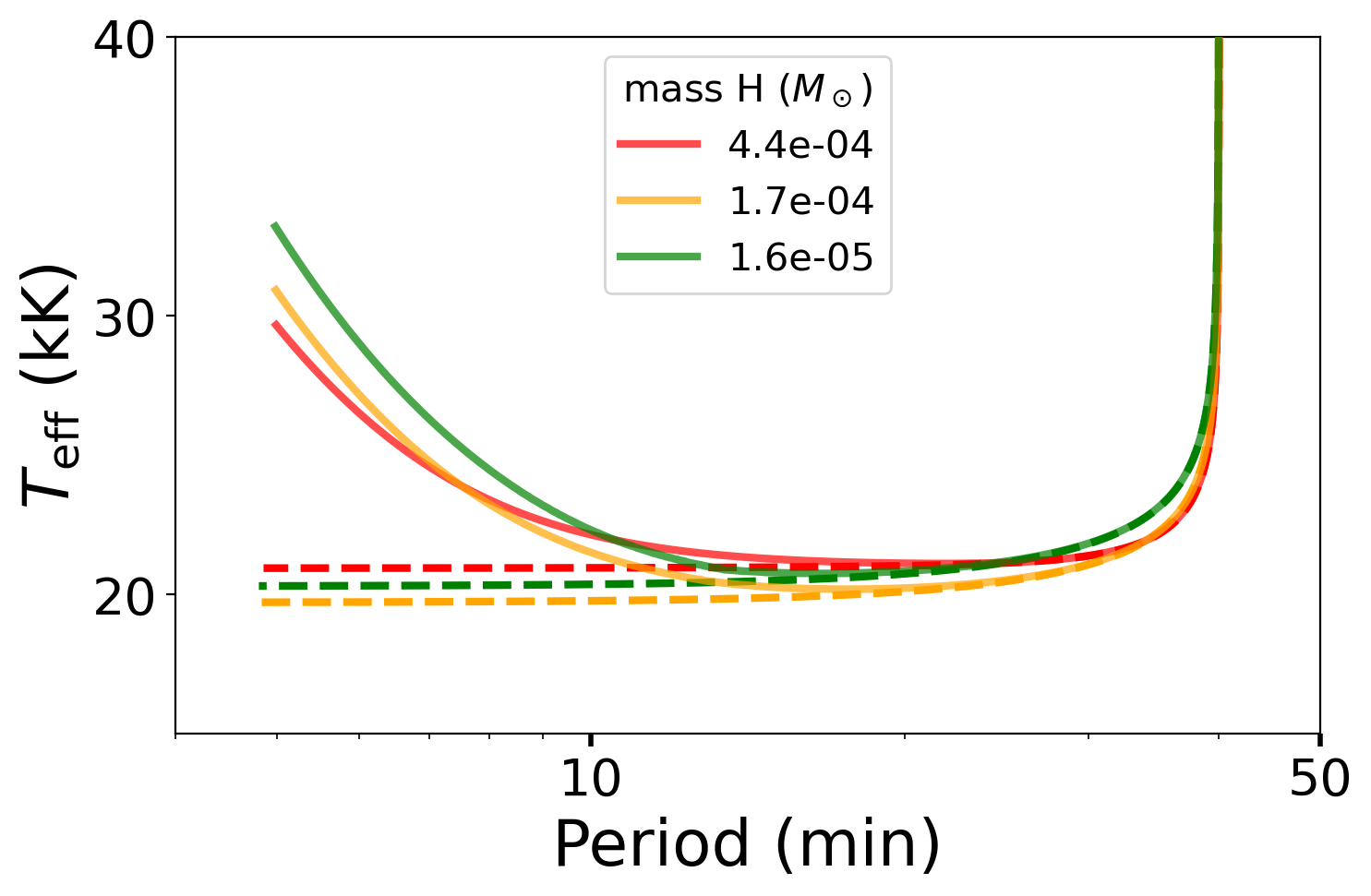}
    \caption{The evolution of effective temperature versus orbital period for several WD models undergoing tidal heating, all of total mass 0.35 $M_\odot$ but with a varying hydrogen envelope mass. Dashed lines correspond to the same models, but with tidal heating turned off. The companion mass is fixed at 0.3 $M_\odot$.}
    \label{fig:H effect}
\end{figure}

He-core WDs can vary in the masses of their H envelopes. We modeled tidal heating for models with a fixed total mass, but varying the mass of the He core and H envelope. Fig. \ref{fig:H effect} demonstrates the evolution of effective temperature for three models of varying H mass, initialized in a binary with the same initial orbital period. Early in the evolution, tidal heating is unimportant but the models exhibit different cooling behavior based on how thick the H envelope is. When tidal heating dominates at short orbital periods, we find that models with a lower mass of H reach higher values of $
T_{\rm eff}$, which is a consistent trend over all our WD models. This trend is not necessarily true when tidal heating is turned off and models continue to cool.

The main cause for this behavior is the changing radius of the WD, as models with more H have a larger radius (for the models in Fig. \ref{fig:H effect}, the radius changes by a factor of $\approx 10\%$).  The radius affects the tidal heating power $L_{\rm tide}$, which scales with the radius $R$, the moment of inertia $I$ and the tidal torque scaling parameter $\hat{f}$ as (see Eq. \ref{f_omega} and \cite{fuller_dynamical_2013} for a derivation)

\begin{equation} \label{power radius}
    L_{\rm tide} \propto \frac{I^{8/7} \hat{f}^{-1/7}}{R^{-31/14}}.
\end{equation}
The values of $I$ and $\hat{f}$ can depend on $R$, but in our models, a changing radius barely changes $I$ and moderately changes $\hat{f}$, which hardly affects $L_{\rm tide}$. At short periods when tidal heating dominates the luminosity, and neglecting the changes in $I$ and $\hat{f}$,

\begin{equation}
    T_{\rm eff}^4 \propto \frac{L_{\rm tide}}{R^2}  \propto \frac{R^{-31/14}}{R^2}  \propto R^{-59/14}.
\end{equation}

In summary, we find that models with a lower mass of H have smaller radii, and reach higher $T_{\rm eff}$ when tidal heating is strong at very short orbital periods. However, the variance in $T_{\rm eff}$ due to varying H mass is $\lesssim \! 10\%$ over a wide range of orbital periods. For the systems that we model, the uncertainties in the total WD mass and initial orbital period of the binary lead to greater uncertainties in $T_{\rm eff}$ (see e.g. Fig. \ref{fig:Teff_grid}). Thus we do not vary the mass of H in our models for our main analysis. Testing our predictions of the effect of H envelope mass on tidal heating would be difficult, both due to its relatively small influence on  $T_{\rm eff}$  and the fact that most WDs do not have measured constraints on their H envelope mass.




\bibliography{main}{}
\bibliographystyle{aasjournal}

\end{document}